\documentclass[conference]{IEEEtran}
\usepackage{blindtext, graphicx}
\usepackage{cite}
\usepackage{multicol}

\usepackage{amssymb}
\usepackage{graphicx}
\usepackage{epstopdf}
\usepackage{url}
\urldef{\mailsa}\path|jiwan.ninglekhu@gmail.com, ram.krishnan@utsa.edu|

\usepackage[]{graphicx}
\usepackage{tabu}
\usepackage{amsmath}
\usepackage{textcomp}
\usepackage{fixltx2e}
\usepackage{amssymb}
\usepackage{ltablex,booktabs}
\usepackage[T1]{fontenc}

\usepackage{tabulary}
\usepackage{xspace}
\usepackage{floatrow}
\usepackage{algorithm}
\usepackage{algorithmic}
\usepackage{hhline}
\makeatletter
\DeclareRobustCommand{\textsupsub}[2]{{%
  \m@th\ensuremath{%
    ^{\mbox{\fontsize\sf@size\z@#1}}%
    _{\mbox{\fontsize\sf@size\z@#2}}%
  }%
}}
\makeatother

\newcommand\tab[1][1.27cm]{\hspace*{#1}}
\newcommand\stab[1][0.635cm]{\hspace*{#1}}
\newcommand\hstab[1][0.3175cm]{\hspace*{#1}}

\newcommand{\U}{USERS}
\newcommand{\R}{ROLES}

\newcommand{\AU}{AU}

\newcommand{\AR}{AR}
\newcommand{\ARH}{ARH}
\newcommand{\AUA}{AUA}

\newcommand{\RH}{RH}

\newcommand{\AATT}{AATT}
\newcommand{\RATT}{RATT}

\newcommand{\ARATT}{ARATT}
\newcommand{\OP}{AOP}

\newcommand{\isauth}{is\_authorized}
\newcommand{\isord}{is\_ordered}

\renewcommand{\it}{\textit}

\newcommand{\q}{\textquotesingle}

\newcommand{\scope}{Scope}

\usepackage[margin=1in,left=1in, includefoot]{geometry}
\usepackage{fancyhdr}
\renewcommand{\headrulewidth}{0pt}

\usepackage{setspace}

\usepackage{booktabs}

\usepackage{enumitem}

\usepackage{cite}

\begin{document}

\title{ARRA: Attribute-Based Role-Role Assignment}

\author{\IEEEauthorblockN{{\textsuperscript{\it{a}}}Jiwan L. Ninglekhu and {\textsuperscript{\it{b}}}Ram Krishnan}
\IEEEauthorblockA{Department of Electrical and Computer Engineering \\
The University of Texas at San Antonio\\
San Antonio, Texas 78249\\
Email: {\textsuperscript{\it{a}}}jiwan.ninglekhu@gmail.com and {\textsuperscript{\it{b}}}ram.krishnan@utsa.edu}
}

\maketitle

\begin{abstract} 
Administrative Role Based Access Control (ARBAC) models specify how to manage user-role assignments (URA), permission-role assignments (PRA), and role-role assignments (RRA). There are many approaches proposed for URA, PRA, and RRA in the literature. In this paper, we propose a model for an attribute-based role- role assignment (ARRA), a novel way to unify prior RRA approaches. We leverage the idea that attributes of various RBAC entities such as admin users and regular roles can be used to administer RRA in a highly flexible manner. We demonstrate that ARRA can express and unify prior RRA models.
\end{abstract}


\IEEEpeerreviewmaketitle

\fancypagestyle{alim}{\fancyhf{}\renewcommand{\headrulewidth}{0pt}\fancyfoot[L]{\footnotesize{\it{\textsupsub{*}{$ $}Article ``AARBAC: Attribute-Based Administration of Role-Based Access Control'' is published at the IEEE CIC 2017 Conference.}}}}
\pagestyle{fancy}
\thispagestyle{alim}
\section{Introduction}
Role-based access control (RBAC)~\cite{sandhu2000nist,sandhu-rbac96} model is well studied in academia\cite{Fuchs2011748} and well adopted in the industry\cite{Oconnor2010nist}. In RBAC, assignment of users and permissions to roles allows users to execute permissions by activating roles. 

Administrative RBAC (ARBAC), which involves in user-role assignment and revocation (URA), permission-role assignment and revocation (PRA), and role-role assignment and revocation (RRA), is a challenging task~\cite{Sandhu97}. ARBAC has been well explored~\cite{Sandhu97, Sandhu99, Oh02,
uarbac, BiswasUni}. All of these works explore URA and PRA. However, only few cover RRA~\cite{Sandhu97, uarbac}. Each of these models use fixed set of properties in making assignment decisions.  For example, 
in RRA97~\cite{Sandhu97}, an admin role is given admin authority over set of roles called role range, where admin role can perform role assignment.

Recently, Attribute-Based Access Control (ABAC) has gained popularity because of its flexibility~\cite{kuhn2010adding, huNISTguide, jin2012unified, bhatt2017abac}. ABAC has also proven its ability to represent different access control models~\cite{jin2012unified}. However, it has rarely been used in administration of RBAC. 

We have previously developed models for attribute-based user-role assignment (AURA) and attribute-based permission-role assignment (ARPA), collectively called the Attribute-Based Administration of Role-Based Access Control (AARBAC)\textsupsub{*}{}. A detailed technical report is presented in~\cite{ninglekhu2017attribute}.
In this paper we present our model for attribute-based role-role assignment (ARRA), a follow-up work on AARBAC~\cite{ninglekhu2017attribute}. ARRA is driven by our objective to design attribute-based model that  allows us to express features from previous RRA models and more. For example, it allows us to express features as a combination of two prior models, and in addition allows us to add new features. Thus,
this work is motivated mainly by two critical factors: (a) 
an objective to build a coherent model, which can be configured to express prior models and more (b) build a unified model that can be referenced for various desirable security properties, which can provide a \it{single} codebase to express prior models and beyond. The contributions of this paper are as follows:
\begin{itemize}
  \item We develop an attribute-based administrative model for role-role
    assignment (\textbf{ARRA}). 
\item We demonstrate that ARRA is capable of expressing prior
  approaches to RRA such as RRA97~\cite{Sandhu97} and UARBAC's RRA~\cite{uarbac}.
\end{itemize}

The remainder of this paper is organized as follows. In
Section~\ref{sec:relatedwork}, we discuss related work.  In
Section~\ref{sec:aarbac}, ARRA model is presented. Section~\ref{sec:URAtranslations} 
presents algorithms that translate prior RRA instances into equivalent ARRA instances. Section~\ref{sec:conclude} concludes the paper.

\section{Related work}\label{sec:relatedwork}

Among many prominent works done for ARBAC~\cite{Sandhu97,Sandhu99,Oh02,BiswasUni,uarbac}, ARBAC97~\cite{Sandhu97} and UARBAC~\cite{uarbac} present role-role assignment (RRA) as part of their model. 
In RRA97, a set of roles in the hierarchy called authority range is given to an admin role for decentralized administration.  An authority range must be an encapsulated range~\cite{Sandhu97}. A user with admin role can perform operations like inserting an edge or creating a role in her authority range. 
 
UARBAC redefines RBAC model with class objects. 
 UARBAC's RRA deals with assigning/revoking role-role on the basis of admin user's authority over individual roles specified by \it{access modes}. 
Class level permissions in UARBAC allows an admin user to conduct an operation over every object of a class. 

In both of these models, the policy for assigning a role to a
role is based on a fixed set of properties of the entities such as admin role, admin user and regular roles, involved in assignment process.

Crampton et. al~\cite{crampton2002administrative} present  models for RRA with administrative scopes. Admin scopes are plausible approach for role hierarchy operations in RBAC. However, admin scopes may not be intuitive to express as a role attribute.
Many literature~\cite{kuhn2010adding,jin2012rabac,rajpoot2015attributes} present benefits of integrating attributes into an RBAC operational model. An operational model deals with making decisions on user's access request for objects. 
There are many works on operational aspects of ABAC~\cite{yuan2005attributed,jin2012unified,huNISTguide,servos2014hgabac}. In contrast, ARRA is an administrative model that uses attributes of RBAC entities for assigning/revoking roles.

\begin{table*}[tbp]
\caption{ARRA Model}
\centering
\label{tab:arra}
\begin{tabu} to 1\textwidth { | X[l] | }
\hline

-- \AU, \OP, \R, \AR\  are finite sets of administrative users, administrative operations such as assign and 

\hspace{7pt} revoke, regular roles and administrative roles, respectively. \\
-- \AUA\ $\subseteq$ \AU\ $\times$ \AR, administrative user to administrative role assignment relation. \\
-- \RH\ $\subseteq$ \R\ $\times$ \R, a partial ordering on the set \R.\\

-- \ARH\ $\subseteq$ \AR\ $\times$ \AR, a partial ordering on the set \AR.\\

-- \AATT, \ARATT, and \RATT\ are finite sets of administrative user attribute functions, administrative role 

\hspace{7pt}attribute functions, and regular role attribute functions, respectively. \\  

-- For each \textit{att} in \AATT\ $\cup$ \ARATT\ $\cup$ \RATT,
\scope(\it{att}) 
is a finite set of atomic values from which the range \\ 
\vspace*{-3ex}
\hspace{7pt}of the 
attribute function \it{att} is derived.
\vspace{1ex}

-- attType : \AATT\ $\cup$ \ARATT\ $\cup$ \RATT\ $\rightarrow$ \{set, atomic\},
which specifies whether the range of a given attribute\\ 
\vspace{-2ex}
\hspace{7pt}is atomic or set valued.\\
-- Each attribute function maps elements in \AU, \AR\ and \R\ to atomic or set values.

  \[
    \forall aatt \in \mbox{\AATT}.\ aatt : \mbox{\AU} \rightarrow \left\{
                \begin{array}{ll}
                  \mbox{\scope}(aatt)\> \mbox{if} \> \mbox{attType}(aatt) = \mbox{atomic}\\
                  2\textsuperscript{\scope(\textit{aatt})}\> \mbox{if}\> \mbox{attType}(aatt) = \mbox{set}\\
                 
                \end{array}
              \right.
  \]
  \[
    \forall aratt \in \mbox{\ARATT}.\ aratt : \mbox{\AR} \rightarrow \left\{
                \begin{array}{ll}
                  \mbox{\scope}(aratt)\> \mbox{if} \> \mbox{attType}(aratt) = \mbox{atomic}\\
                  2\textsuperscript{\scope(\textit{aratt})}\> \mbox{if}\> \mbox{attType}(aratt) = \mbox{set}\\
                 
                \end{array}
              \right.
  \]
\[
  \forall ratt \in \mbox{\RATT}.\ ratt : \mbox{\R} \rightarrow \left\{
            \begin{array}{ll}
                \mbox{\scope}(ratt)\> \mbox{if} \> \mbox{attType}(ratt) = \mbox{atomic}\\
                 2\textsuperscript{\scope(\textit{ratt})}\> \mbox{if}\> \mbox{attType}(ratt) = \mbox{set}\\
               
               \end{array}
             \right.
\]
-- \isord\ :  \AATT\ $\cup$ \ARATT\ $\cup$ \RATT\ $\rightarrow$ \{True, False\}, specifies
if the scope is ordered for each of the\\
\hspace{7pt}attributes.\\
-- For each \it{att} $\in$ \AATT\ $\cup$ \ARATT\ $\cup$ \RATT,

\hspace{7pt}if \isord(\it{att}) = True, 
H\textsubscript{\it{att}} $\subseteq$ \scope(\it{att}) $\times$ \scope(\it{att}),
a partially ordered attribute hierarchy, and H\textsubscript{\it{att}} $\neq$
$\phi$,

\hspace{7pt}else, 
if \isord(\it{att}) = False, 
H\textsubscript{\it{att}} = $\phi$\\
\hspace{3pt}
(For some \it{att} $\in$ \AATT\ $\cup$ \ARATT\ $\cup$ \RATT\
for which attType(\it{att}) = set and \isord(\it{att}) = True,
if
$\{a,b\}$, 

\hspace{7pt}$\{c,d\}$
$\in$ 2\textsuperscript{\scope(\textit{att})}
(where $a, b, c, d \in$ \scope($att$)), 
we infer
$\{a,b\}$ $\geq$ $\{c,d\}$ if
$(a,c)$, $(a,d)$, $(b,c)$, $(b,d)$

\hspace{7pt}$\in$ H\textsupsub{*}{\it{att}}.)
\vspace{7pt}\\
\hhline{|=|}

ARRA model allows an administrator to perform an operation on a single role
or a set of roles at a time. The authorization rule for performing an operation on a single role is as follows:\\

For each \it{op} in \OP, {\textbf{\isauth}}\textbf{R\textsubscript{\it{op}}}(\it{au}: \AU, \it{r\textsubscript1} :
\R, \it{r\textsubscript2} : \R) specifies if the admin user \it{au} is allowed to perform
the operation \it{op} (e.g. assign, revoke, etc.) between the
regular roles \it{r\textsubscript1} and the role \it{r\textsubscript2}. Note that assigning a role \it{r\textsubscript1} to \it{r\textsubscript2} makes \it{r\textsubscript1} junior to \it{r\textsubscript2}. This rule is written as a logical expression using attributes of admin
user \it{au}, admin role, \it{ar}, and regular role, \it{r}.
\vspace{1em}\\
\hline
The authorization rule for performing an operation on a set of users is as follows:\\

For each \it{op} in \OP, {\textbf{\isauth}}\textbf{R\textsubscript{\it{op}}}(\it{au}: \AU, $\chi$ :
2\textsuperscript{\R}, \it{r} : \R) specifies if the admin user \it{au} is allowed to
perform the operation \it{op} (e.g. assign, revoke, etc.) between the roles
in the set $\chi$ and the role \it{r}. 

Here $\chi$ is a set of roles that can be specified using a set-builder
notation, whose rule is written using role attributes.
\vspace{1em}
\\
\hline
\end{tabu}
\end{table*}

\section{ARRA Model}\label{sec:aarbac}
In this section, we present our approach for attribute-based role-role assignment (ARRA). Inspired by prior RRA models we have included attributes for admin users, admin roles and regular roles. Based on which, decisions on assigning a role to a role is made. 
%

Table~\ref{tab:arra} presents formal ARRA model. The entities involved in ARRA comprise of admin users (\AU), regular roles (\R) and their
hierarchy (\RH), admin roles (\AR) and their hierarchy (\ARH), admin user to admin role relation (\AUA) and admin operations (\OP). 

In ARRA, admin user in \AU\ wants to perform admin operation such as assign or revoke from \OP\ using attributes of entities in the model. We have developed admin users attribute functions (\AATT) and admin roles attribute functions (\ARATT). Based on the need we have observed, we have also introduced regular roles attribute functions (\RATT). 
We will later see that we need attributes from different entities in representing properties of RRA97 and UARBAC's RRA in ARRA.   

The attribute functions or \emph{simply} attributes are defined as mapping from its domain such as \AU\ or \AR\ to its range. Range of an attribute \it{att} can be atomic or set valued, which is specified by function attType. It is derived from set of scope or atomic values denoted by \scope(\it{att}). Furthermore, the scope of an attribute can be either ordered or unordered, which is given by a function called \isord(\it{att}). If it is ordered, we must specify that attribute's hierarchy denoted by H\textsubscript{\it{att}} on its scope \scope(\it{att}). H\textsubscript{\it{att}} is a partial ordering on \scope(\it{att}). Note that even in the case of a set valued attribute \it{att}, the
hierarchy H\textsubscript{\it{att}} is specified on \scope(\it{att}) instead of
2\textsuperscript{\scope(\it{att})}. We infer the ordering between two set values given an ordering on atomic values. 
H\textsupsub{*}{\it{att}} in Table~\ref{tab:arra}
denotes the reflexive transitive closure of
H\textsubscript{\it{att}}.

In ARRA, there are two ways by which an admin user can select a set of regular roles for assignment to a target regular role. The first way allows an admin user to select a single role and a target role, and perform an admin operation like assign or revoke. The second way allows an admin user to select a set of regular roles, the target role and perform similar operation on those roles. In the latter case, the selection criteria for the set
of regular roles can be expressed using a set-builder notation whose rule
is based on the regular role attributes. For example, 
{\isauth}R\textsubscript{\textbf{assign}}(\textbf{au},
\{\it{r\textsubscript1} | \it{r\textsubscript1} $\in$ \R\ $\wedge$ \textbf{Lead} $\in$ \it{roleTitle}(\it{r\textsubscript1})\},
\textbf{r\textsubscript2}) 
would specify a policy for an admin user \textbf{au} that selects the set of all the roles with role title \textbf{Lead} in order to assign those roles to a role \textbf{r\textsubscript2}. Assigning a role \it{r\textsubscript1} to role \it{r\textsubscript2} make role \it{r\textsubscript1} junior to \it{r\textsubscript2}, i.e., it adds an entry <\it{r\textsubscript2, r\textsubscript1}> in \RH. It is referred to as edge insetion in RRA97. 

Authorization rule is specified as a logical expression on
the attributes of admin users, admin roles, and that of regular roles considered for assignment.

\section{Mapping Prior RRA Models in ARRA}\label{sec:URAtranslations}

In this section, we demonstrate that ARRA can intuitively simulate
the features of prior RRA models. In particular, we have developed
concrete algorithms that can convert any instance of RRA97 and UARBAC's RRA into their corresponding equivalent ARRA instances. Due to space limitation, we have included explicit example instances for each of these prior models, their manual translation into corresponding ARRA instances, and detail description in as separate article~\cite{ninglekhu2017arra} 

\subsection{RRA97 in ARRA}
Map\textsubscript{RRA97} (Algorithm~\ref{alg:rra97}) maps any RRA97
instance into an equivalent ARRA instance. Labels {97} and {A} represent sets and functions from RRA97 and ARRA, respectively. Map\textsubscript{RRA97} takes an instance of RRA97 as input.
In particular, input consists of \U\textsuperscript{{97}},
\R\textsuperscript{{97}}, \AR\textsuperscript{{97}}, \AUA\textsuperscript{{97}}, \RH\textsuperscript{{97}}, \ARH\textsuperscript{{97}}, and \it{can\_modify}\textsuperscript{{97}}. The
\it{can\_modify} instruction covers operations for inserting an edge, deleting an edge, creating a role and deleting a role. ARRA model can simulate inserting an edge and deleting an edge. However, creating and deleting roles are beyond the scope of current ARRA model.

Output from Map\textsubscript{RRA97} algorithm is 
an equivalent ARRA instance with  \AU\textsuperscript{{A}}, \OP\textsuperscript{{A}}, \R\textsuperscript{{A}}, \AUA\textsuperscript{{A}}, \RH\textsuperscript{{A}}, \ARH\textsuperscript{{A}}, \ARATT\textsuperscript{{A}}, For each
attribute \it{att} $\in$ \ARATT\textsuperscript{{A}},
\scope\textsuperscript{{A}}(\it{att}), attType\textsuperscript{{A}}(\it{att}),
\isord\textsuperscript{{A}}(\it{att}) and H\textsuperscript{{A}}\textsubscript{\it{att}}, 
For each admin role \it{ar} $\in$ \AR\textsuperscript{{A}} and for each \it{att} $\in$ \ARATT\textsuperscript{{A}}, \it{att}(\it{ar}), 
Authorization rule for assign
(auth\_assign), and Authorization rule for revoke (auth\_revoke). 

As indicated in Map\textsubscript{RRA97}, Step 1 maps sets from RRA97 to ARRA sets. 
In Step 2, admin role attribute is
expressed. 
\AATT\ and \RATT\ are left empty as there is no use case for these attributes in translating RRA97. Admin role attribute \it{authRange} captures the mapping between an authority range as defined in RRA97, and an admin role. It is a set valued and unordered attribute with its scope as a transitive closure of regular role hierarchy, \RH\textsuperscript{+}. In ARRA, we use a symbolic representation of an authority range with end points a and b, as
(a, b). However, whenever we need roles present in (a, b), we denote the set of roles with $\lceil$a, b$\rceil$, i.e., $\lceil$a, b$\rceil$ = \{\it{r} $\vert$ \it{r} $\in$ \R\textsuperscript{{{A}}} $\wedge$ a < \it{r} < b\}.

In Step 3, we construct an authorization rule for \it{insert edge} operation from \it{can-modify}\textsuperscript{{97}}, which is 
equivalent to {\isauth}R\textsubscript{\textbf{insertEdge}}(\it{au} : \AU\textsuperscript{{A}}, \it{r\textsubscript1} :
\R\textsuperscript{{A}}, \it{r\textsubscript2} : \R\textsuperscript{{A}}) in ARRA. This formula asserts assignment conditions that preserves encapsulation for admin roles' authority ranges. Similarly, In Step 4, authorization rule equivalent to deleting an edge from \it{can\_modify}\textsuperscript{{97}} is expressed.
\vspace{0.06cm}
\floatname{algorithm}{Algorithm}
\begin{algorithm}
\caption{Map\textsubscript{RRA97}}
\label{alg:rra97}
\begin{algorithmic} [1] 
\begin{spacing}{1.22}

\item[]\hspace{-17pt}\textbf{Input:} RRA97 instance
\item[]\hspace{-17pt}\textbf{Output:} ARRA instance 
\item[\textbf{Step 1:}]  \  /* Map basic sets and functions in AURA */
\item[] a. \AU\textsuperscript{{A}} $\leftarrow$ \U\textsuperscript{{97}} \\
\item[] b. \OP\textsuperscript{{A}} $\leftarrow$ \{\textbf{insertEdge, deleteEdge}\} 
\item[] c. \R\textsuperscript{{A}} $\leftarrow$ \R\textsuperscript{{97}} ; 
 \AUA\textsuperscript{{A}} $\leftarrow$ \AUA\textsuperscript{{U}} 
\item[] b. \RH\textsuperscript{{A}} $\leftarrow$ \RH\textsuperscript{{97}} 
\item[\textbf{Step 2:}]   \stab /* Map attribute functions in AURA */
\item[] a. \AATT\textsuperscript{{A}} $\leftarrow$ \{\} ; \RATT\textsuperscript{{A}} $\leftarrow$ \{\}
\item[] b. \ARATT\textsuperscript{{A}} $\leftarrow$ \{\it{authRange}\} 
\item[] c. \scope\textsuperscript{{A}}(\it{authRange}) = \RH\textsuperscript{+}  \textsuperscript{{A}} 
\item[] d. attType\textsuperscript{{A}}(\it{authRange}) = set 
\item[] e. \isord\textsuperscript{{A}}(\it{authRange}) = False ; H\textsupsub{{A}}{\it{authRange}} = $\phi$
\item[] f. For each \it{ar} $\in$ \AR\textsuperscript{{A}}, \it{authRange}(\it{ar}) = $\phi$  
\item[] g. For each (\it{ar}, (\it{r\textsubscript{i}, r\textsubscript{j}})) $\in$ \it{can\_modify}\textsuperscript{97},
\item[] \stab \it{authRange}(\it{ar}) = \it{authRange}(\it{ar}) $\cup$ (\it{r\textsubscript{i}, r\textsubscript{j}})
\item[\textbf{Step 3:}] \stab /* Construct assign rule in AURA */
\item[] a. assign\_formula = $\phi$ 
\item[] b. For each (\it{ar}, (\it{r\textsubscript{i}, r\textsubscript{j}})) $\in$ \it{can\_modify}\textsuperscript{97}, 
\item[] \stab assign\_formula\textquotesingle\ = assign\_formula $\vee$ \\
($\exists$(\it{au}, \it{ar\textsubscript1}) $\in$ \AUA\textsuperscript{{A}}, $\exists$(s, t) $\in$ \it{authRange}(\it{ar\textsubscript1}). \\
\it{r\textsubscript1}, \it{r\textsubscript2} $\in$ $\lceil$s, t$\rceil$) 
$\wedge$
(($\exists$(m, n), (m\q, n\q) $\in$ \\
$\bigcup\limits_{ar\textsubscript2 \in\ \textrm{\AR}}$ \it{authRange}(\it{ar\textsubscript2}). 
\it{r\textsubscript1}, \it{r\textsubscript2} $\in$ $\lceil$m, n$\rceil$ $\wedge$ ($\lceil$m\q, n\q$\rceil$ $\subset$ $\lceil$m, n$\rceil$ $\rightarrow$ \it{r\textsubscript1}, \it{r\textsubscript2} $\notin$ (m\q, n\q)))
$\vee$ 
($\exists$(x,y) $\in$ $\bigcup\limits_{ar\textsubscript3 \in\ \textrm{\AR}}$ \it{authRange}(\it{ar\textsubscript3}). ((\it{r\textsubscript1} = y $\wedge$ \it{r\textsubscript2} > x) $\vee$ (\it{r\textsubscript2} = x $\wedge$ \\
\it{r\textsubscript1} < y)) $\wedge$ ($\forall$p $\in$ $\lceil$x, y$\rceil$ $\wedge$ $\forall$q $\notin$ $\lceil$x, y$\rceil$. (<q, p> $\in$ (\RH\textsuperscript{{A}} $\cup$ <\it{r\textsubscript2}, \it{r\textsubscript1}>)* $\leftrightarrow$ <q, y> $\in$ (\RH\textsuperscript{{A}} $\cup$ <\it{r\textsubscript2}, \it{r\textsubscript1}>)*) $\wedge$ (<p, q> $\in$ (\RH\textsuperscript{{A}} $\cup$ <\it{r\textsubscript2}, \it{r\textsubscript1}>)*  $\leftrightarrow$ <x, q> $\in$ (\RH\textsuperscript{{A}} $\cup$ <\it{r\textsubscript2}, \it{r\textsubscript1}>)*))))
\item[] c. auth\_assign = {\isauth}R\textsubscript{\textbf{insertEdge}}(\it{au} : \AU\textsuperscript{{A}}, \\
\item[] \stab \it{r\textsubscript1} : \R\textsuperscript{{A}}, \it{r\textsubscript2} : \R\textsuperscript{{A}}) $\equiv$ assign\_formula\textquotesingle
\item[\textbf{Step 4:}] \stab /* Construct revoke rule for AURA */
\item[] a. revoke\_formula = $\phi$
\item[] b. For each (\it{ar\textsubscript1}, (\it{r\textsubscript{1}, r\textsubscript{2}})) $\in$ \it{can\_modify}\textsuperscript{97},
\item[] \stab revoke\_formula\textquotesingle\ = revoke\_formula $\vee$ \\
\item[] $\exists$(\it{au}, \textit{ar}) $\in$ \AUA\textsuperscript{{A}} $\wedge$ $\exists$(x, y) $\in$ \it{authAR}(\it{ar}). 
\it{r\textsubscript1}, \it{r\textsubscript2} \\
$\in$ (x, y) $\wedge$ $\exists$(\it{r\textsubscript1}, \it{r\textsubscript2}) $\notin$ $\bigcup\limits_{ar \in\ \textrm{\AR}}$ \it{authRange}(\it{ar})
\item[] c. auth\_revoke = {\isauth}R\textsubscript{\textbf{deleteEdge}}(\it{au} : \AU\textsuperscript{{A}}, \\
\item[] \stab \it{r\textsubscript1} : \R\textsuperscript{{A}}, \it{r\textsubscript2} : \R\textsuperscript{{A}}) $\equiv$ revoke\_formula\textquotesingle
\vspace{-13pt}
\end{spacing}
\end{algorithmic}
\end{algorithm}

\floatname{algorithm}{Algorithm}

\begin{algorithm}
\caption{Map\textsubscript{RRA-UARBAC}}
\label{algo:rra-uarbac}
\begin{algorithmic} [1] 
\begin{spacing}{1.005}
\item[]\hspace{-17pt}\textbf{Input:} Instance of RRA in UARBAC 
\item[]\hspace{-17pt}\textbf{Output:} ARRA instance 
\item[\textbf{Step 1:}] \ \  /* Map basic sets and functions in AURA */
\item[] a. \AU\textsuperscript{{A}} $\leftarrow$ \it{\U}\textsuperscript{{U}} ; \OP\textsuperscript{{A}} $\leftarrow$ \{\textbf{assign, revoke}\}
\item[] b. \R\textsuperscript{{A}} $\leftarrow$ \it{\R}\textsuperscript{{U}} ;  \AUA\textsuperscript{{A}} = $\phi$
\item[] c. \RH\textsuperscript{{A}} $\leftarrow$ \it{\RH}\textsuperscript{{U}} 

\item[\textbf{Step 2:}]   \stab /* Map attribute functions in AURA */
\item[] a. \AATT\textsuperscript{{A}} $\leftarrow$ \{\it{grantAuth, empowerAuth, adminAuth, 
\item[] \stab roleClassAuth}\}, \ARATT\textsuperscript{{A}} = \{\}, \RATT\textsuperscript{{A}} = \{\}
\item[] b. \scope\textsuperscript{{A}}(\it{grantAuth}) = \R\textsuperscript{{A}},  
\item[] c. attType\textsuperscript{{A}}(\it{grantAuth}) = set 
\item[] d. \isord\textsuperscript{{A}}(\it{grantAuth}) = True, 
H\textsuperscript{{A}}\textsubscript{\it{grantAuth}} = \RH\textsuperscript{{A}}
\item[] e. For each \it{u\textsubscript{}} in \AU\textsuperscript{{U}}, \it{grantAuth}(\it{u\textsubscript{}}) = $\phi$ 
\item[] f. For each \it{u} in \it{\U}\textsuperscript{{U}} and 
\item[] \ \ \  for each [\textsf{role}, \it{r\textsubscript{}}, \textsf{grant}] $\in$ authorized\_perms\textsuperscript{{U}}[\it{u}],
\item[] \tab \it{grantAuth}(\it{u\textsubscript{}})\textquotesingle\ = \it{grantAuth}(\it{u\textsubscript{}}) $\cup$ \it{r\textsubscript{}}
\item[] g. \scope\textsuperscript{{A}}(\it{empowerAuth}) = \R\textsuperscript{{A}},  
\item[] h. attType\textsuperscript{{A}}(\it{empowerAuth}) = set 
\item[] i. \isord\textsuperscript{{A}}(\it{empowerAuth}) = True, 
\item[] j. H\textsuperscript{{A}}\textsubscript{\it{empowerAuth}} = \RH\textsuperscript{{A}}
\item[] k. For each \it{u\textsubscript{}} in \AU\textsuperscript{{U}}, \it{empowerAuth}(\it{u\textsubscript{}}) = $\phi$ 
\item[] l. For each \it{u} in \it{\U}\textsuperscript{{U}} and 
\item[] \stab for each [\textsf{role}, \it{r\textsubscript{}}, \textsf{empower}] 
\item[] \tab \stab  $\in$ authorized\_perms\textsuperscript{{U}}[\it{u}],
\item[] \ \ \ \  \it{empowerAuth}(\it{u\textsubscript{}})\textquotesingle\ = \it{empowerAuth}(\it{u\textsubscript{}}) $\cup$ \it{r\textsubscript{}}
\item[] m. \scope\textsuperscript{{A}}(\it{adminAuth}) = \R\textsuperscript{{A}}
\item[] n. attType\textsuperscript{{A}}(\it{adminAuth}) = set 
\item[] o. \isord\textsuperscript{{A}}(\it{adminAuth}) = True, 
\item[] p. H\textsuperscript{{A}}\textsubscript{\it{adminAuth}} = \RH\textsuperscript{{A}} 
\item[] q. For each \it{u\textsubscript{}} in \U\textsuperscript{{A}}, \it{adminAuth}(\it{u}) = $\phi$ 
\item[] r. For each \it{u} in \it{\U}\textsuperscript{{U}} and
\item[] \ \ \ for each [\textsf{role}, \it{r\textsubscript{}}, \textsf{admin}] $\in$ authorized\_perms\textsuperscript{{U}}[\it{u}], 
\item[] \tab \it{adminAuth}(\it{u\textsubscript{}})\textquotesingle\ = \it{adminAuth}(\it{u\textsubscript{}}) $\cup$ \it{r\textsubscript}
\item[] s. \scope\textsuperscript{{A}}(\it{roleClassAuth}) = \it{AM}\textsuperscript{{U}}(\textsf{role}) 
\item[] t. attType\textsuperscript{{A}}(\it{roleClassAuth}) = set 
\item[] u. \isord\textsuperscript{{A}}(\it{roleClassAuth}) = False 
\item[] v. H\textsuperscript{{A}}\textsubscript{\it{roleClassAuth}} = $\phi$ 
\item[] w. For each \it{u\textsubscript{}} in \U\textsuperscript{{A}}, \it{roleClassAuth}(\it{u}) = $\phi$ 
\item[] x. For each \it{u} in \it{\U}\textsuperscript{{U}} 
\item[] \stab for each [\it{c, am}] $\in$ authorized\_perms\textsuperscript{{U}}[\it{u}], 
\item[]\tab \it{roleClassAuth}(\it{u})\q = \it{roleClassAuth}(\it{u}) $\cup$ \it{am} 
\item[\textbf{Step 3:}] \stab /* Construct assign rule in AURA */
\item[] a. assign\_formula = 
\hspace*{0.3cm}(\it{r\textsubscript{1}} $\in$ \it{grantAuth}(\it{au\textsubscript{}}) $\wedge$ \\ 
\hspace*{0.3cm}\it{r\textsubscript2} $\in$ \it{empowerAuth}(\it{au\textsubscript{}})) $\vee$ (\it{r\textsubscript{1}} $\in$ \it{grantAuth}(\it{au\textsubscript{}}) $\wedge$ \\ 
\hspace*{0.3cm}\textsf{empower} $\in$ \it{roleClassAuth}(\it{au\textsubscript{}})) $\vee$ \\
\hspace*{0.3cm}(\textsf{grant} $\in$ \it{roleClassAuth}(\it{au\textsubscript{}}) $\wedge$ \\
\hspace*{0.3cm}\it{r\textsubscript2} $\in$ \it{empowerAuth}(\it{au\textsubscript{}})) $\vee$\\
\hspace*{0.3cm}(\textsf{grant} $\in$ \it{roleClassAuth}(\it{au\textsubscript{}}) $\wedge$ \\
\hspace*{0.3cm}\textsf{empower} $\in$ \it{roleClassAuth}(\it{au\textsubscript{}}))

\vspace{-0.5cm}
\end{spacing}
\end{algorithmic}
\end{algorithm}


\floatname{algorithm}{Continuation of Algorithm}
\setcounter{algorithm}{1}
\begin{algorithm}
\caption{Map\textsubscript{RRA-UARBAC}}
\begin{algorithmic} [1] 
\begin{spacing}{0.85}
\item[]

\item[] b. auth\_assign = 
\item[] \ \ \ \ {\isauth}U\textsubscript{\textbf{assign}}(\it{au\textsubscript{}} : \AU\textsuperscript{{A}}, \it{r\textsubscript1} : \R\textsuperscript{{A}}, \it{r\textsubscript2} : \R\textsuperscript{{A}}) $\equiv$   assign\_formula
\item[\textbf{Step 4:}] \stab /* Construct revoke rule for AURA */
\item[] a. revoke\_formula = 
(\it{r\textsubscript{1}} $\in$ \it{grantAuth}(\it{au\textsubscript{}}) $\wedge$  \\ 
\hspace*{0.3cm}\it{r\textsubscript2} $\in$ \it{empowerAuth}(\it{au\textsubscript{}})) $\vee$ \it{r\textsubscript{1}} $\in$ \it{adminAuth}(\it{au\textsubscript{}}) $\vee$\\
 \hspace*{0.3cm}\it{r\textsubscript{2}} $\in$ \it{adminAuth}(\it{au\textsubscript{}}) $\vee$
\textsf{admin} $\in$ \it{roleClassAuth}(\it{au})
\item[] b. auth\_revoke = 
\item[]  \ \ \ \ {\isauth}U\textsubscript{\textbf{revoke}}(\it{au\textsubscript1} : \AU\textsuperscript{{A}}, \it{r\textsubscript1} : \R\textsuperscript{{A}}, \it{r\textsubscript2} : \R\textsuperscript{{A}}) $\equiv$ revoke\_formula
\vspace{-0.3cm}
\end{spacing}
\end{algorithmic}
\end{algorithm}

\subsection{UARBAC's RRA in ARRA}
\label{sec:rra-uarbac}
Algorithm~\ref{algo:rra-uarbac} presents
Map\textsubscript{RRA-UARBAC} that maps any instance of UARBAC's RRA ~\cite{uarbac} to its equivalent ARRA instance. 
Labels {U} and {A} represent sets and functions from UARBAC and ARRA, respectively.
Input to Map\textsubscript{RRA-UARBAC} consists of 
\it{C}\textsuperscript{{U}}, \it{\U}\textsuperscript{{U}}, \it{\R}\textsuperscript{{U}}, \it{P}\textsuperscript{{U}},
\it{\RH}\textsuperscript{{U}},
\it{AM}\textsuperscript{{U}}(\textsf{role}), For each \it{u} $\in$ \it{\U}\textsuperscript{{U}},
authorized\_perms\textsuperscript{{U}}[\it{u}], and
For every \it{r\textsubscript1}, \it{r\textsubscript2} $\in$ \it{\R}\textsuperscript{{U}}, grant operation
grantRoleToRole(\it{r\textsubscript1, r\textsubscript2}) will be true if the granter has either {[\textsf{role}, \it{r\textsubscript2}, \textsf{empower}]} and  {[\textsf{role}, \it{r\textsubscript1}, \textsf{grant}]} \textbf{or},
{[\textsf{role}, \it{r\textsubscript{2}}, \textsf{empower}]} and {[\textsf{role}, \textsf{grant}]} \textbf{or},
{[\textsf{role}, \textsf{empower}]} and {[\textsf{role}, \it{r\textsubscript1}, \textsf{grant}]} \textbf{or}, {[\textsf{user}, \textsf{empower}]} and {[\textsf{role}, \textsf{grant}]} permissions towards roles. 

 For each \it{r\textsubscript1}, \it{r\textsubscript2} $\in$ \it{\R}\textsuperscript{{U}}, revokeRoleFromUser(\it{r\textsubscript1, r\textsubscript2}) is true if the granter has either  
{[\textsf{role}, \it{r\textsubscript{2}}, \textsf{empower}] and [\textsf{role}, \it{r\textsubscript1}, \textsf{grant}]} \textbf{or}, 
{[\textsf{role}, \it{r\textsubscript1}, \textsf{admin}]} \textbf{or}, 
 {[\textsf{role}, \it{r\textsubscript2}, \textsf{admin}]} \textbf{or},
 {[\textsf{role}, \textsf{admin}]} permission on roles.

Map\textsubscript{RRA-UARBAC} yields an ARRA instance consisting of \AU\textsuperscript{{A}}, \OP\textsuperscript{{A}},
\R\textsuperscript{{A}}, \AR\textsuperscript{{A}}, \AUA\textsuperscript{{A}}, \RH\textsuperscript{{A}}, \ARH\textsuperscript{{A}}, \AATT\textsuperscript{{A}}, For each attribute
\it{att} $\in$  \AATT\textsuperscript{{A}},
\scope\textsuperscript{{A}}(\it{att}), attType\textsuperscript{{A}}(\it{att}),
\isord\textsuperscript{{A}}(\it{att}) and H\textsuperscript{{A}}\textsubscript{\it{att}}, For each user
\it{au} $\in$ \AU\textsuperscript{{A}} and for each \it{att} $\in$ \AATT\textsuperscript{{A}}, \it{att}(\it{au}), Authorization rule for assign
(auth\_assign), and Authorization rule for revoke (auth\_revoke).

In Step 1 primary sets from 
UARBAC's RRA are mapped to AURA equivalent sets. In Step 2, 
admin user attributes are defined. In UARBAC model, assignment decisions of role to role are based on the admin user's \emph{access modes} such as \textsf{grant}, \textsf{empower} and \textsf{admin} towards regular roles. It treats each entity in the RBAC system such as files, roles and users as objects. We define admin user attributes \it{grantAuth, empowerAuth} and \it{adminAuth} to capture an admin user's
\it{access modes} on regular role, respectively. For example, \it{grantAuth} yields a set of roles over which, an admin user has \textsf{grant} access mode.  Attribute \it{roleClassAuth} specifies the form of class level \it{access mode} an admin user has on \textsf{role} class. Class level \it{access mode} on \textsf{role} class for instance gives an admin user authority with particular \it{access mode} over all the available roles. 

In Step 3, an authorization rule equivalent to
grantRoleToRole(\it{r\textsubscript1, r\textsubscript2}) from UARBAC is expressed. In ARRA it is represented by {\isauth}R\textsubscript{\textbf{assign}}(\textit{au\textsubscript} : \AU\textsuperscript{{A}}, \it{r\textsubscript1} : \R\textsuperscript{{A}}, \textit{r\textsubscript2} : \R\textsuperscript{A}), which is equal to assign\_formula.
Similarly, in Step 4, authorization rule for revoke equivalent to revokeRoleFromRole(\it{r\textsubscript1, 
r\textsubscript2}) is constructed using attributes of an admin user.  
\vspace{-0.11cm}

\subsection{An Example Instance for ARRA with role attributes}
Previous ARRA model simulation examples did not include regular role attribute. In this section, we present a simple yet plausible example that demonstrates a use case for role attributes in assigning a role to a role. 

There are two admin users, \textbf{Sam} and \textbf{Tom}, and regular roles \textbf{IT Director, Development Manager, Quality Manager, Marketing Manager, Finance Manager, Support Engineer} and \textbf{System Analyst}. There can be many other roles in an organization. We have considered few roles enough to illustrate our case. Admin user attribute \it{dept} captures admin user's authority over set of departments. Although, it is likely that departments have hierarchy in practice, we have not included department hierarchy for simplicity. There are three departments in the organization, namely, \textbf{Operations, Account} and \textbf{IT}. \textbf{Sam} has admin authority over all the departments while \textbf{Tom} has authority over \textbf{IT} department only. There is a regular role attribute \it{dept} for mapping regular roles to their departments.

Authorization condition is that if an admin user \it{au} with admin role \it{ar} was given authority over a department \it{d} and if roles \it{r\textsubscript1} and \it{r\textsubscript2} belonged to the same department \it{d} then \it{r\textsubscript1} can be assigned to \it{r\textsubscript2}. 
\\
\underline{Sets}\\
-- \AU\ = \{\textbf{Sam, Tom}\}, \OP\ = \{\textbf{assign, revoke}\} \\
-- \R\ = \{\textbf{IT Director, Development Mgr., Quality} \\
\stab \textbf{Mgr., Marketing Mgr., Finance Mgr., \\
\stab Support Engineer, System Analyst}\} \\
-- \AR\ = \{\textbf{}\}, \AUA\ = \{\}, \RH\ = \{\},  \ARH\ = \{\} \\
\underline{Attributes and functions} \\
-- \AATT\ = \{\it{dept}\}, \\
-- \scope(\it{dept}) = \{\textbf{Operations, Account, IT}\}, \\ 
\hstab attType(\it{dept}) = set, \isord(\it{dept}) = False, \\
\hstab H\textsubscript{\it{dept}} = $\phi$\\ 
-- \it{dept}(\textbf{Sam}) = \{\textbf{Operations, Account, IT}\}, \\
\hstab \it{dept}(\textbf{Tom}) = \{\textbf{IT}\}  \\
-- \ARATT\ = \{\}, \RATT\ = \{\it{dept, level}\} \\
-- \scope(\it{dept}) = \{\textbf{Operations, Account, IT}\}, \\ 
\hstab attType(\it{dept}) = atomic, \\
\hstab \isord(\it{dept}) = False, H\textsubscript{\it{dept}} = $\phi$ \\
-- \it{dept}(\textbf{IT Director}) = \textbf{IT},\\
\hstab \it{dept}(\textbf{Marketing Mgr.}) = \textbf{Operations}, \\
\hstab \it{dept}(\textbf{QA Mgr.}) = \textbf{IT}, \it{dept}(\textbf{Development Mgr.}) = \textbf{IT}, \\
%
%
\underline{Authorization functions}\\
\noindent
-- {\isauth}R\textsubscript{\textbf{assign}}(\it{au} : \AU, \it{r\textsubscript1} : \R, \it{r\textsubscript2} : \R) $\equiv$ 
$\exists$\it{d} $\in$ \scope(\it{dept}). \it{dept}(\it{au}) = \it{dept}(\it{r\textsubscript1}) = \it{dept}(\it{r\textsubscript2}) 
\newline
-- {\isauth}R\textsubscript{\textbf{revoke}}(\it{au} : \AU, \it{r\textsubscript1} : \R, \it{r\textsubscript2} : \R) $\equiv$ {\isauth}R\textsubscript{\textbf{assign}}(\it{au} : \AU, \it{r\textsubscript1} : \R, \it{r\textsubscript2} : \R)
\vspace{-0.2cm}

\section{Conclusion}\label{sec:conclude}
In this paper, we presented ARRA, a model for attribute based role-role assignment. A design motivation behind ARRA model was to make it enough to express prior RRA models. In particular, we took RRA97 and UARBAC's RRA model as foundation. For these models, we have presented Map\textsubscript{RRA97} and Map\textsubscript{RRA-UARBAC} algorithms, which map any instances of these models to equivalent ARRA instances, respectively. 

ARRA model can not only express existing RRA models but also has a capability to do more. For instance, prior models included in our research did not fit the notion of regular role attributes. However, we acknowledge that it is an essential attribute in making role assignment decisions to say the least. Motivated by this notion, we presented a simple example scenario where we made use of regular role attributes for role-role assignment.

\section*{Acknowledgment}
This work is partially supported by NSF grants CNS-1423481 and CNS-1553696.

\ifCLASSOPTIONcaptionsoff
  \newpage
\fi

\bibliographystyle{IEEEtran}
\bibliography{bib/References}

\end{document}